# CVD Graphene Contacts for Lateral Heterostructure MoS2 Field Effect Transistors

Daniel S. Schneider[1,2], Leonardo Lucchesi[3], Eros Reato[1,2], Zhenyu Wang[4], Agata Piacentini[1,2], Jens Bolten[1], Damiano Marian[3], Enrique G. Marin[5], Aleksandra Radenovic[4], Zhenxing Wang[1], Gianluca Fiori[3], Andras Kis[4], Giuseppe Iannaccone[3], Daniel Neumaier[1,6] and Max C. Lemme[1,2*]

[1]Advanced Microelectronic Center Aachen (AMICA), AMO GmbH, 52074 Aachen, Germany.

[2]RWTH Aachen University, 52074 Aachen, Germany.

[3]University of Pisa, 56122 Pisa, Italy.

[4]École Polytechnique Fédérale de Lausanne (EPFL), CH-1015, Lausanne, Switzerland.

[5]University of Granada, 18070 Granada, Spain.

[6]University of Wuppertal, 42285 Wuppertal, Germany

[*]Email: lemme@amo.de / Phone: (+49) 241 8867 201

ABSTRACT

Intensive research is carried out on two-dimensional materials, in particular molybdenum disulfide, towards high-performance transistors for integrated circuits[1]. Fabricating transistors with ohmic contacts is challenging due to the high Schottky barrier that severely limits the transistors' performance. Graphene-based heterostructures can be used in addition to, or as a substitute for unsuitable metals. We present lateral heterostructure transistors made of scalable chemical vapor-deposited molybdenum disulfide and chemical vapor-deposited graphene with low contact resistances of about 9 k$\Omega\cdot\mu$m and high on/off current ratios of $10^8$. We also present a theoretical model calibrated on our experiments showing further potential for scaling transistors and contact areas into the few nanometers range and the possibility of a substantial performance enhancement by means of layer optimizations that would make transistors promising for use in future logic circuits.

Two-dimensional (2D) semiconducting materials from the group of transition metal dichalcogenides (TMDCs) are promising for aggressively scaled transistors for next-generation integrated circuits that are largely unaffected by short-channel effects[2–5]. By now, stable wafer-based deposition techniques have been achieved for the most studied TMDC molybdenum disulfide ($MoS_2$)[6–9]. Also, $MoS_2$ has been demonstrated as a suitable channel material for n-type field-effect transistors (FETs) with high performance[10–12] and has been successfully used in circuits[13–15]. Low-voltage and low-power applications require transistors that not only exhibit sufficiently high mobility and high current on/off ratio but also low contact resistance between the metal electrodes and the 2D channel. Direct contacting of $MoS_2$ with metals can lead to the formation of Schottky barriers and Fermi-level pinning at the interfaces. The results are undesirably high contact resistances and therefore limited carrier injection and, ultimately, reduced device performance[16,17]. For this reason, various contact methods were evaluated. The use of $MoS_2$ phase-transformed from semiconducting 2H into metallic 1T has been proposed as a contact region[18]. However, it should be noted that stable 1T contact regions have not yet been reproducibly demonstrated with chemical vapor deposited (CVD)-grown $MoS_2$. Others have demonstrated that ultra-high vacuum-deposited Au contacts can also lead to very low contact resistance[12,19,20]. However, gold is typically not suitable for monolithic integration due to the lack of silicon CMOS compatibility. A further approach is chloride molecular doping, which leads to a significant reduction of contact resistance but is also unstable over time[21]. Recently, McLellan et al. [22] presented a stable doping process that can lead to very low contact resistance through aluminum oxide ($AlO_x$) capping, but also induces strong n-doping and cannot easily be limited to the contact regions[22]. Another approach to achieving very low contact resistances to 2D materials is to use semimetals like bismuth and antimony[23,24], although they are relatively scarce, which may limit their long-term use as contact metals.[25] Graphene, on the other hand, has excellent electrical conductivity and is chemically stable, making it a promising candidate for low-resistance contacts to $MoS_2$ [26–34]. Several reports have already proven the versatility of graphene as a contact due to the tunability of its work function by electrostatic doping, which minimizes the Schottky barrier height at the graphene-semiconductor interface[26,32,35–37]. Using the semimetal graphene as contacts provides an atomically sharp interface without dangling bonds, so the Fermi-level pinning[38] at the contact interface to TMDCs can be prevented[33]. The mechanical strength, flexibility, transparency, and thermal stability of 2D materials make them highly desirable for flexible electronics applications[39]. Here, device bendability is a significant concern. Graphene contacts can offer an excellent alternative to traditional metals, which are often the limiting factor in high strain levels. Although the advantages of graphene are apparent, the graphene sheet resistance and the metal-graphene contact resistance contribute to the total resistance in addition to the graphene-TMDC contact resistance and, therefore, must be co-optimized in the contact design and fabrication. Furthermore, the employment of graphene contacts required an additional metallization layer for electrical characterization due to the inherent thinness of graphene as an electrode. An intrinsic limitation of these contacts is the high van der Waals gap, which is, in some cases, higher than the metal-$MoS_2$ gap[40]. This issue can, however, be addressed by achieving a perfect edge contact between the metal and the graphene [41].

In this work, we experimentally demonstrate a scalable technique for low contact resistance to MoS$_2$ using CVD-grown SLG. We further show that metal one-dimensional edge contacts between single-layer graphene (SLG) and nickel (Ni) are a suitable method for achieving low contact resistances[42,43]. The results are corroborated through simulations that explore the scalability potential of further contact resistance optimization based on this approach.

Results

Material Characterization

Confocal Raman measurements were performed on the SLG/MoS$_2$ lateral heterostructure (LH) (**Figure 1**a) to analyze the transferred 2D films. A Raman map with 400 points taken after the transfer of SLG on SiO$_2$ in **Figure 1**b shows the 2D (2692 cm$^{-1}$) and G (1589 cm$^{-1}$) modes of SLG. A defect peak around 1350 cm$^{-1}$ was not detected, indicating high quality graphene after the transfer. The thickness of MoS$_2$, as measured by atomic force microscopy in a prior growth process, is approximately 0.7 nm and aligns with the Raman data confirming its monolayer structure.[9]

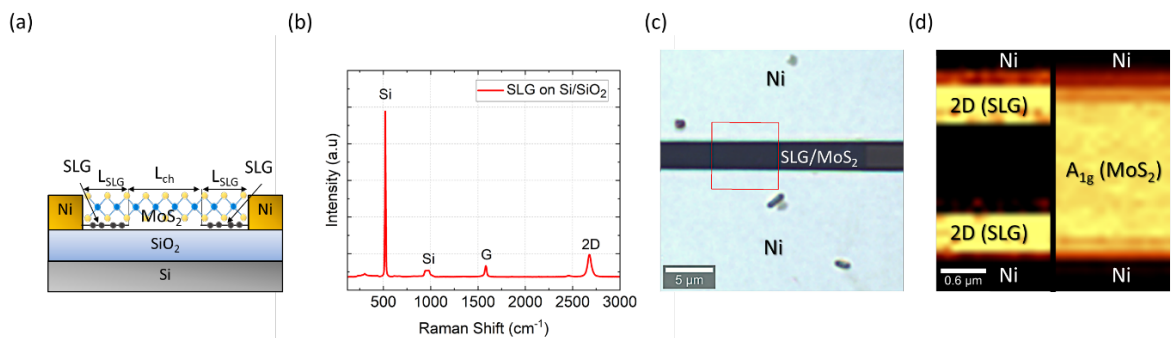

**Figure 1: Overview of the sample design and 2D-layers arrangement.** (a) Device schematic of LH-FETs. $L_{SLG}$ describes the length of the SLG/MoS$_2$ heterostructure and $L_{ch}$ defines the channel of the transistor. (b) Raman spectrum of SLG on Si/SiO$_2$ before the transfer of MoS$_2$. (c) Optical microscope image of a LH-FET. The Raman area scan shown in (d) was performed in the area marked by the red box. (d) A spatially resolved Raman map shows the intensity of the 2D mode of SLG (left) and the intensity of the A$_{1g}$ mode of MoS$_2$ (right). Dark areas indicate not present mode while brighter areas indicate a stronger intensity.

A spatially resolved Raman map of a µm-scale FET was performed (red box in **Figure 1**c). The strong intensity of the A$_{1g}$ mode of MoS$_2$ in **Figure 1**d confirms that the MoS$_2$ channel uniformly covers the entire region between the Ni contacts, while SLG is in contact with the Ni and the MoS$_2$.

Device Characteristics

The contact resistance $R_c$ of Ni edge contacts to SLG is typically in the range of only a few hundred Ω µm and the sheet resistance $R_s$ of SLG is ~1 kΩ/square[43] (Supplementary Fig. 4), more than 100x lower than that of MoS$_2$.

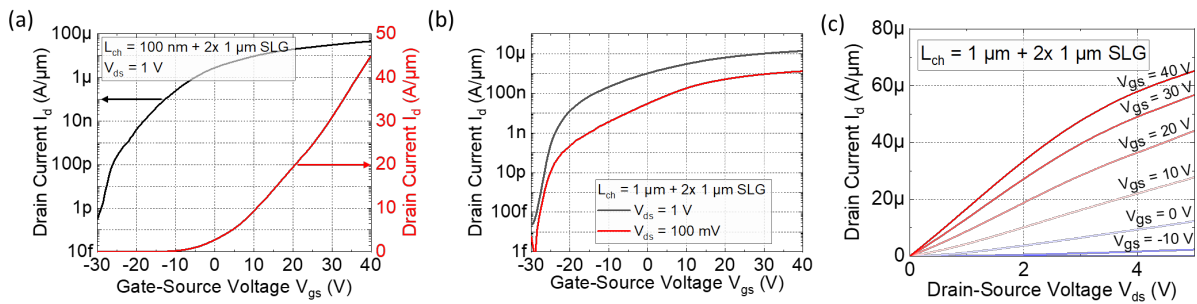

**Figure 2: Transistor performance characteristics of the devices.** (a) Transfer characteristic of a MoS$_2$ LH-FET with $L_{ch}$ = 100 nm for $V_{ds}$ = 1 V in log scale (black line) and linear scale (red line). (b) Transfer

characteristic of a LH-FET with $L_{ch}$ = 1 μm for $V_{ds}$ = 1 V (black line) and $V_{ds}$ = 100 mV (red line) and (c) its corresponding output curves.

Figure 2a shows the transfer characteristic for $V_{ds}$ = 1 V with measured maximum on current of $I_{on}/W$ = 43 μA/μm for a device with $L_{ch}$ = 100 nm and $V_{gs}$ = 40 V.

The transfer characteristics of a LH-FET with $L_{ch}$ = 1 μm is shown in **Figure 2**b. This device reached a high current on/off ratio of more than $10^8$ with a low off-current of order 10 fA/μm at $V_{gs}$ = -30 V (a graph with a wider $V_{gs}$ sweep is shown in Supplementary Fig. 3). The transfer curves show a kink at $V_{gs}$ = -20 V, which can be attributed to acceptor-like interface states in S/D regions[44]. The output curves of the 1 μm-long LH-FET in **Figure 2**c demonstrate an ohmic behavior of the drain currents, indicating the suitable contacting scheme with graphene. The drain current's saturation range is limited by the relatively thick gate oxide (90 nm $SiO_2$), which hinders the build-up of a strong electrostatic potential across the channel. The on-currents of the device increase continuously as the channel length decreases with approximately a $1/\sqrt{L_{ch}}$ relationship, as shown in **Figure 3**a for $V_{ds}$ = 100 mV (red) and $V_{ds}$ = 1V (black) at $V_{gs}$ = 40V. The transmission line method (TLM) was used to extract a contact resistance from the total resistance

$$R_{total} = 2 \cdot R_c + R_{ch} \quad (1)$$

where $R_{ch}$ is the channel resistance. **Figure 3**b shows the linear fit of $R_{total}$ for different $L_{ch}$, and the point at which the line intersects the Y-axis corresponds to a value of $2 \cdot R_c$. Here, a $R_c$ = 9 ±2 kΩ μm was extracted for the LH-FET, which is more than one order of magnitude lower than reference MoS$_2$ FETs with pure Ni side contacts (Supplementary Fig. 5 and a comparison of transfer characteristics of LH-FETs and Ni contacted FETs in Supplementary Fig. 6). For both types of devices, a sheet resistance of $R_s$ ~60 kΩ /square was extracted. To demonstrate the scalability of our approach, we conducted measurements on 10 LH-FETs for each channel length (Supplementary Fig. 8), revealing minimal device-to-device variability. A benchmarking plot of literature data compares contact resistances of multilayer CVD and exfoliated MoS$_2$ FETs contacted with graphene (**Figure 3**c). The comparison has been made among graphene-contacted MoS$_2$-FETs to emphasize the importance of this scalable approach, which still has a margin of improvement in the quality of the materials. Even though flake-based MoS$_2$ and graphene have previously achieved lower contact resistances, this work stands out by utilizing scalable materials and demonstrating the lowest reported contact resistance on a CVD monolayer MoS$_2$ channel, narrowing the gap with the superiority with flakes-based materials.

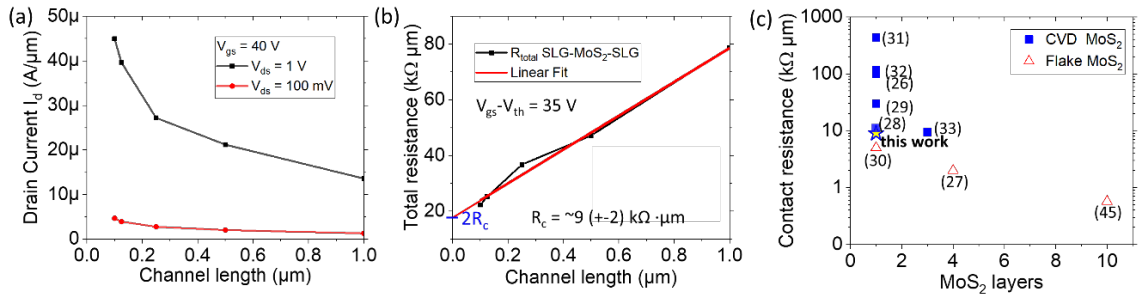

**Figure 3: Additional performance data and benchmarking of the contact resistances.** (a) $I_{on}$ is plotted against $L_{ch}$ for $V_{ds}$ = 1 V (black line) and $V_{ds}$ = 100 mV (red line) in double-logarithmic scale. (b) Total device resistance vs channel length of different LH-FETs and the extracted contact resistance by TLM.

(c) Benchmarking plot of contact resistance of graphene-contacted MoS$_2$ FETs [26–30,32,33,45] as a function of MoS$_2$ layers (blue: CVD MoS$_2$; red: exfoliated MoS$_2$).

Device Simulation

To better understand device operation and the potential of downscaling to nanometer size, we performed a multiscale simulation of the LH-FET.

In our model, we describe the top contact as a ladder of resistors and current generators, as shown in **Figure 4**a. The horizontal (in-plane) resistors have a resistance proportional to the layers' sheet resistances and the vertical current generators provide the vertical current per unit area due to ballistic transport, which is nonlinearly dependent on the electrochemical potentials of the two nodes and on the vertical electrostatic potential profile. Since we can define local electrochemical potentials for each layer $\mu_{Gr}(x)$ and $\mu_{MoS_2}(x)$ on the horizontal direction, the nonlinear current generators are described through a modified Landauer formula where, in order to use measurable quantities, we use the in-layer applied potentials defined as

$$V_i(x) = -(1/e)\mu_i(x) \qquad (2)$$

instead of the chemical potentials:

$$I_v(x) = I_0 \cdot \int T\left(E, \phi\left(V_{Gr}(x), V_{MoS_2}(x)\right)\right)\left(f\left(E + V_{MoS_2}(x)\right) - f\left(E + V_{Gr}(x)\right)\right) dE. \qquad (3)$$

Assuming that the electrostatic potential varies smoothly in the horizontal direction, we can compute the electrostatic potential for every horizontal position $x$ through a vertical 1D Poisson simulation dependent the on the gate voltage $V_{gs}$ and the applied potentials. In general, the interlayer transmission coefficient $T(E, x)$ depends on the detailed shape of the electrostatic potential in the vertical direction. Assuming that the difference between the electrochemical potential of the two layers is small, we can approximate $T(E, x)$ as a $T(E + e\phi(x))$, where $\phi(x)$ is the average of the electrostatic potentials of the two layers for the same horizontal coordinate x: $\phi(x) = (\phi_{Gr}(x) + \phi_{MoS_2}(x))/2$ (4)

This corresponds to considering that the shape of the barrier in the vertical direction is negligibly dependent on the electrochemical potential of the layers and that the transmission coefficient in the vertical direction is only affected by the average shift of the barrier. We computed $T(E)$ through a multiscale approach considering two infinite layers at equilibrium following the procedure described in the methods section.

The solution of the Poisson equations enables us at the same time to use the aforementioned approximation to shift $T(E)$ to $T(E + e\phi(x))$, and to compute the carrier densities in the two layers, which in turn affect the sheet resistances

$$R_{sh,a}(x) = \frac{1}{[e\mu_{c,a} n_{s,a}(x)]} \quad (5)$$

, where $n_{s,a}(x)$ is the majority carrier density and $\mu_c$ is the carrier mobility in the layer denoted by $a$, where $(a = Gr, \text{MoS}_2)$.

The effect of the gate voltage $V_{gs}$ is therefore entirely embodied in $T(E + \phi(x))$ and $R_{sh,1,2}(x)$. The model parameters are the materials' mobilities $\mu_{c,Gr}$, $\mu_{c,MoS2}$, and the corresponding doping densities $n_{Gr}$, $n_{MoS_2}$. We estimated $n_{Gr}$ on the SiO2 substrate to be $4.5 \times 10^{12} \text{cm}^{-2}$ from the position of the charge neutral point in an analogous structure in Ref. [43], while for the MoS2 we assume it to be $2 \times 10^{12} cm^{-2}$. Using the estimated doping densities, we could use carrier densities obtained from Poisson simulations to estimate $\mu_{c,Gr}$ and $\mu_{c,MoS2}$ respectively from the sheet resistance of Ref. [43] and the channel sheet resistance obtained from the TLM depicted in Supplementary Fig. 7 by inverting

$$R_{sh} = \frac{1}{[e\mu_c n_s]} \quad (6)$$

for MoS2 we estimated it to be $\mu_{c,MoS2} \approx 1 \text{ cm}^2\text{V}^{-1}\text{s}^{-1}$ for sample 1 and $\mu_{c,MoS2} \approx 7 \text{ cm}^2\text{V}^{-1}\text{s}^{-1}$ for sample 2, and for graphene we estimated it to be $\mu_{c,Gr2} \approx 629 \text{ cm}^2\text{V}^{-1}\text{s}^{-1}$. We also introduced a single fitting parameter: the interface quality factor $\eta$, defined as

$$\eta = \frac{I_0}{I_{0,Landauer}} \quad (7)$$

which represents the missing knowledge on the actual distance between the planes and the overall interface quality.

We cast our model in a non-linear transmission line model structure, where we solved a system of differential equations describing horizontal transport with source terms representing vertical transport[43]. Our model allowed us to obtain the applied potentials $V_{Gr,MoS_2}(x)$ and the horizontal currents $I_{Gr,MoS_2}(x)$ profiles in the two layers. From these current profiles, we could extract two main quantities describing the performance of the top contact *i)* contact resistance ($R_c$) and *ii)* transfer length ($\lambda_T$). The former is the main figure of merit of a contact. With our calculation we confirm the experimental result by obtaining $R_c = 9 \, k\Omega \cdot \mu m$ for the estimated mobility of sample 2. As for the latter, which represents the characteristic length over which the current goes from one layer to the other, and therefore the minimum length for the top contact, we obtain $\lambda_T = 27 \, nm$, showing potential for contact length scaling and integration. Low resistance is achieved by inserting the semimetal graphene between the metal and the 2D semiconductor. As demonstrated in Shen et al.[23], this partially avoids the formation of the hybridized states in the gap of the semiconductor that lead to the Fermi pinning phenomenon. The relatively low Schottky barrier between graphene and MoS$_2$ of ~0.5 eV can be further reduced electrostatically via the gate voltage, since graphene is a semimetal, and we cannot assume it to have an infinite density of states[26,35,36,47]. Furthermore, the gate voltage introduces a large carrier density in the contact, further improving conductivity. The use of an advanced fabrication process also ensures a clean heterostructure with a low density of impurities and defects.

The main confirmation of the validity of our model is the correct scaling of contact resistance $R_c$ with $\mu_{c,MoS2}$. As we can see in **Figure 4**d, by increasing $\mu_{c,MoS2}$ while keeping every other parameter constant implies a nonlinear decrease in $R_c$. Our mobility scaling can describe $R_c$ in two samples with different mobilities by using the same value for $\eta$. This suggests that $\eta$ is a general parameter, depending only on materials choice and interface quality, not on the quality of the materials. Therefore, obtaining the value of $\eta$ for one value of $\mu_{c,MoS2}$ allowed us to extrapolate the contact resistance for contacts made with higher quality materials, assessing the potential of the Ni-graphene/MoS₂ contact for future practical applications. We can see that even if the contact would have been made of materials with record mobilities (~10000 cm²V⁻¹s⁻¹) for graphene [44] and ~200 cm²V⁻¹s⁻¹ for MoS₂ [10]), with the current interface quality $\eta = 0.17$ the minimum contact resistance would have been $R_c \sim 2\ k\Omega \cdot \mu m$ with a transfer length of $\lambda_T \sim 80\ nm$. For an ideal interface quality $\eta = 1$, we can estimate the minimum achievable contact resistance to be $R_c \sim 0.5\ k\Omega \cdot \mu m$ with $\lambda_T \sim 35\ nm$. **Figure 4**d also shows that interface quality is even more important than the 2D-materials mobilities since the contact resistance quickly saturates because of the effective resistance of the vertical interface.

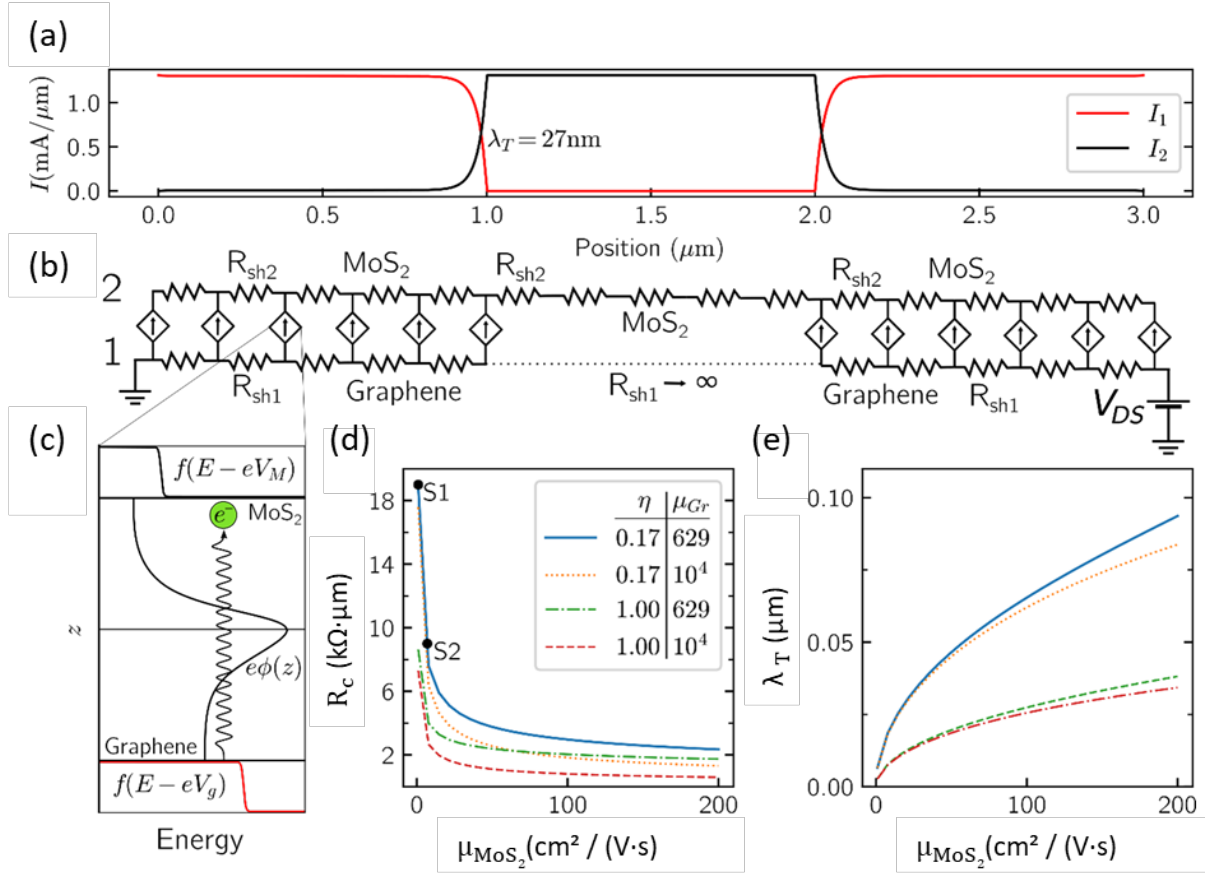

Figure 4: **Multiscale simulation of transport.** (a) Simulated horizontal current for each layer in the whole device for typical device operation conditions. Horizontal current transfer between the graphene layer (1) and the MoS2 layer (2) shows presence of vertical carrier transport over a few transfer lengths $\lambda_T$. (b) Circuit-like schematization of device, a non-linear Transmission Line Model setup. Vertical ballistic transport between the two resistive layers is represented by nonlinear current generators (c) Depiction of vertical ballistic transport. Carriers propagate through the interlayer barrier $\phi(z)$ from the local Fermi distributions generated by local electrostatic potentials $V_{1,2}(x)$. (d) Simulated dependence of contact resistance $R_c$ on MoS2 mobility $\mu_{MoS_2}$ for different interface quality $\eta$ and graphene mobility $\mu_{Gr}$. Our simulation correctly predicts the difference in contact resistance between two samples (S1 and S2) with different $\mu_{MoS_2}$. (e) Simulated transmission length $\lambda_T$ dependence on $\mu_{MoS_2}$ for different $\eta$ and $\mu_{Gr}$ (same legend as (d) ). $\lambda_T$ does not depend strongly on $\mu_{Gr}$.

In summary, we have experimentally demonstrated lateral SLG/MoS2 heterostructures based on scalable materials, with low contact resistances down to ~9 kΩ μm at current ON/OFF ratios of $10^8$. The proposed theoretical model, calibrated with experiments, shows a charge transfer length down to 27 nm, indicating the scaling potential of the SLG approach for ultra-scaled 2D FETs. Furthermore, our model shows that TMDCs with higher mobility and an optimized interface can lead to a very promising contact resistance of 0.5 kΩ µm. Here, direct growth processes or cleaner transfers of large area grown 2D materials are necessary to improve the interfaces of heterostructures and thus lower contact resistance in future scalable devices.

METHODS

MoS$_2$ deposition

A continuous single-layer MoS$_2$ film was grown by metal-organic chemical vapor deposition (MOCVD) on 2" sapphire wafer using molybdenum hexacarbonyl (Mo(CO)$_6$) and hydrogen disulfide (H$_2$S) precursors[9]. Extensive characterization of the material has been performed in reference [9], including transmission electron microscopy (TEM) to assess the quality of the material.

Material Characterization

Confocal Raman and PL measurements were performed with a laser wavelength of 532 nm and a power of 1 mW on MoS$_2$ in detail, both on its growth substrate sapphire and after being transferred onto 90 nm silicon oxide on silicon substrates (Supplementary Fig. 1). PL measurements of MoS$_2$ and Raman measurements of SLG were conducted with a 300 lines/mm grating and Raman measurements of MoS$_2$ with a 1800 lines/mm grating. The step height of ~0.7 nm of MoS$_2$ was measured by atomic force microscopy.

Device Fabrication

Commercially available CVD grown SLG on copper (Cu) was transferred onto a 90 nm SiO$_2$/Si substrate with pre-patterned alignment marker via PMMA supported wet transfer. Electron beam lithography (EBL) and Oxygen (O$_2$) plasma reactive ion etching (RIE) were used to pattern the SLG contact areas. CVD-MoS$_2$ was then transferred onto the entire chip by wet transfer[48]. FET channels were defined by EBL and subsequent RIE using a gas mixture of tetrafluoromethane (CF$_4$) and O$_2$. Finally, self-aligned sputtered Ni edge contacts to SLG were defined by EBL and a subsequent CF$_4$/O$_2$ plasma RIE process using the same resist mask. The back-gated FETs with different channel lengths from 100 nm to 1 μm were used to determine the contact resistance by Transfer-Line-Method (TLM). The relatively large device channel width of 100 μm was used to compensate single material defects in the 2D layers or residues caused by the transfers[49]. A sketch of the fabrication process of lateral heterostructure (LH)-FETs is shown in Figure S2. Although graphene and MoS$_2$ are vertically stacked, the term 'lateral heterostructure 'refers to the lateral transistor configuration wherein the MoS$_2$ channel can be modulated between two graphene contacts.

Computational Methods

The device model required a correct description of the physics and the two largely different length scales in the horizontal direction (~1 μm) and in the vertical direction (~1 nm). For this reason, we assume diffusive transport in the horizontal direction and ballistic transport in the vertical direction across the van-der-Waals gap of the heterojunction. These two very different transport regimes were described with a single multiscale model. As for the ballistic transport, we compute the vertical transmission coefficient between graphene and $MoS_2$ following the procedure detailed in Ref. [50], which consists in performing *i)* a density functional theory (DFT) simulations of the infinite graphene-$MoS_2$ heterostructure, *ii)* a transformation of the DFT Hamiltonian into the basis of maximally localized Wannier function (MLWF) using proper projection in order to clearly identify the top and bottom flake, i.e. $MoS_2$ and graphene, *iii)* non-equilibrium Green's function (NEGF) simulation to compute the vertical transmission, creating a proper MLWF Hamiltonian with monolayer and bilayer regions[50]. DFT calculations have been carried out using Quantum Espresso suite[51]. We have considered a supercell consisting of 5 × 5 graphene and 4 × 4 $MoS_2$ elementary cells, applying 3% of strain on the graphene and no strain on $MoS_2$, with an interlayer distance of 3.4 A. We use GGA-PBE pseudopotentials and grimme-D2 correction to consider van der Waals forces. Calculations are performed on a 3 × 3 × 1 grid. The Hamiltonian in terms of the MLWF has been obtained exploiting Wannier90 code[52] projecting on the $p_z$ orbital of each C atom and on the three *sp2* orbitals every two C atoms while on the *d*-orbitals for the Mo and on the *s*- and the *p*-orbitals for the S atoms. The transmission coefficient has been obtained using NanoTCAD ViDES[53]

DATA AVAILABILITY STATEMENT

Data sets generated during the current study are available from the corresponding authors on reasonable request.

ACKNOWLEDGMENTS

We acknowledge the European Union's Horizon 2020 research and innovation program under the grant agreements QUEFORMAL (829035) and Graphene Flagship (881603), the German Research Foundation (DFG) projects MOSTFLEX (407080863), ULTIMOS$_2$ (LE 2440/8-1) and INST 221/96-1, as well as the German Ministry of Education and Research (BMBF) projects NeuroTec II (16ME0399, 16ME0400) and NeuroSys (03ZU1106AA).

AUTHOR CONTRIBUTIONS

The manuscript was written through the contributions of all authors. All authors have given approval to the final version of the manuscript.

COMPETING INTERESTS

The Authors declare no Competing Financial or Non-Financial Interests.

ASSOCIATED CONTENT

Supplementary material is available online or from the author.

AUTHOR INFORMATION

Corresponding Author

*Lemme@amo.de

FIGURE CAPTIONS

**Figure *1*: Overview of the sample design and 2D-layers arrangement.** (a) Device schematic of LH-FETs. $L_{SLG}$ describes the length of the SLG/MoS$_2$ heterostructure and $L_{ch}$ defines the channel of the transistor. (b) Raman spectrum of SLG on Si/SiO$_2$ before the transfer of MoS$_2$. (c) Optical microscope image of a LH-FET. The Raman area scan shown in (d) was performed in the area marked by the red box.

(d) A spatially resolved Raman map shows the intensity of the 2D mode of SLG (left) and the intensity of the $A_{1g}$ mode of $MoS_2$ (right). Dark areas indicate not present mode while brighter areas indicate a stronger intensity.

**Figure 2: Transistor performance characteristics of the devices.** (a) Transfer characteristic of a $MoS_2$ LH-FET with $L_{ch}$ = 100 nm for $V_{ds}$ = 1 V in log scale (black line) and linear scale (red line). (b) Transfer characteristic of a LH-FET with $L_{ch}$ = 1 µm for $V_{ds}$ = 1 V (black line) and $V_{ds}$ = 100 mV (red line) and (c) its corresponding output curves.

**Figure 3: Additional performance data and benchmarking of the contact resistances.** (a) $I_{on}$ is plotted against $L_{ch}$ for $V_{ds}$ = 1 V (black line) and $V_{ds}$ = 100 mV (red line) in double-logarithmic scale. (b) Total device resistance vs channel length of different LH-FETs and the extracted contact resistance by TLM. (c) Benchmarking plot of contact resistance of graphene-contacted $MoS_2$ FETs [26–30,32,33,45] as a function of $MoS_2$ layers (blue: CVD $MoS_2$; red: exfoliated $MoS_2$).

**Figure 4: Multiscale simulation of transport.** (a) Simulated horizontal current for each layer in the whole device for typical device operation conditions. Horizontal current transfer between the graphene layer (1) and the $MoS_2$ layer (2) shows presence of vertical carrier transport over a few transfer lengths $\lambda_T$. (b) Circuit-like schematization of device, a non-linear Transmission Line Model setup. Vertical ballistic transport between the two resistive layers is represented by nonlinear current generators (c) Depiction of vertical ballistic transport. Carriers propagate through the interlayer barrier $\phi(z)$ from the local Fermi distributions generated by local electrostatic potentials $V_{1,2}(x)$. (d) Simulated dependence of contact resistance $R_c$ on $MoS_2$ mobility $\mu_{MoS_2}$ for different interface quality $\eta$ and graphene mobility $\mu_{Gr}$. Our simulation correctly predicts the difference in contact resistance between two samples (S1 and S2) with different $\mu_{MoS_2}$. (e) Simulated transmission length $\lambda_T$ dependence on $\mu_{MoS_2}$ for different $\eta$ and $\mu_{Gr}$ (same legend as (d) ). $\lambda_T$ does not depend strongly on $\mu_{Gr}$.

# CVD Graphene Contacts for Lateral Heterostructure MoS$_2$ Field Effect Transistors

*Daniel S. Schneider[1,2], Leonardo Lucchesi[3], Eros Reato[1,2], Zhenyu Wang[4], Agata Piacentini[1,2], Jens Bolten[1], Damiano Marian[3], Enrique G. Marin[5], Aleksandra Radenovic[4], Zhenxing Wang[1], Gianluca Fiori[3], Andras Kis[4], Giuseppe Iannaccone[3], Daniel Neumaier[1,6] and Max C. Lemme[1,2\*]*

[1]Advanced Microelectronic Center Aachen (AMICA), AMO GmbH, 52074 Aachen, Germany.

[2]RWTH Aachen University, 52074 Aachen, Germany.

[3]University of Pisa, 56122 Pisa, Italy.

[4]École Polytechnique Fédérale de Lausanne (EPFL), CH-1015, Lausanne, Switzerland.

[5]University of Granada, 18070 Granada, Spain.

[6]University of Wuppertal, 42285 Wuppertal, Germany

[\*]Email: lemme@amo.de / Phone: (+49) 241 8867 201

Confocal Raman and photoluminescence (PL) measurements were performed with a laser wavelength of 532 nm and a power of 1 mW on $MoS_2$ in detail, both on its growth substrate sapphire and after being transferred onto 90 nm silicon oxide on silicon substrates (Supplementary Fig. 1). PL and Raman measurements were conducted with a 300 g/mm and an 1800 g/mm grating, respectively. The Raman spectrum in Supplementary Fig. 1a showing the $A_{1g}$ (404 cm$^{-1}$) and $E_{2g}$ (384 cm$^{-1}$) modes of $MoS_2$ before and $A_{1g}$ (406 cm$^{-1}$) and $E_{2g}$ (383 cm$^{-1}$) after the transfer onto $SiO_2$, respectively. The distance of the $A_{1g}$ and $E_{2g}$ of 20 cm$^{-1}$ indicating a monolayer $MoS_2$ film thickness on the growth substrate[1]. The PL spectrum in Supplementary Fig. 1b show a strong A-exciton peak emission at 1.86 eV (red curve) and thereby emphasizes a high film quality. The suppression of the peak (black curve) after the transfer is due to the different reflections of the $Si/SiO_2$ substrate[2].

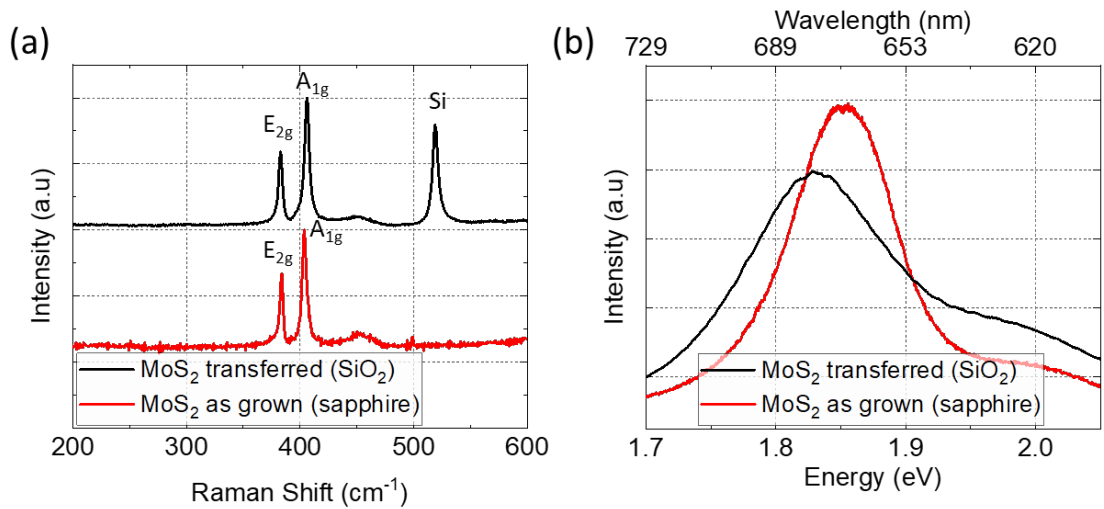

**Supplementary Figure 1:** (a) Confocal Raman and (b) PL measurements of the as grown $MoS_2$ monolayer (red line) on sapphire and transferred onto $SiO_2$ substrate (black line).

A schematic device cross-section is shown in Supplementary Fig. 2. All devices were designed with a channel width of 100 μm. The length of the graphene contacts ($L_{SLG}$) are 1 μm and are not included in the actual transistor channel lengths ($L_{ch}$). $L_{ch}$ is defined by the gap between the SLG contacts, vary between 100 nm (Fig.1b) and 1 μm.

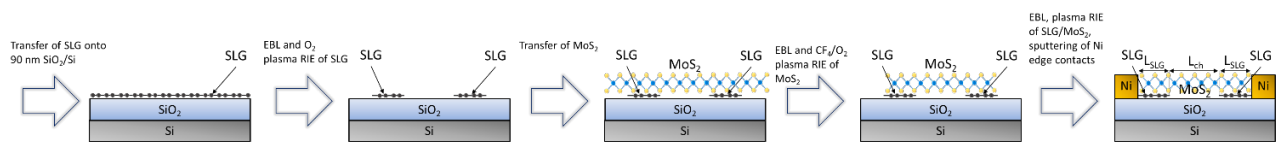

**Supplementary Figure 2**: Fabrication process and device schematic of LH-FETs. Pre-structured graphene contacts at the bottom serve as low ohmic contacts to $MoS_2$. The 2D-heterostructure is contacted by nickel edge contacts. The contact length of graphene is denoted as L and the $MoS_2$ channel in between as $L_{ch}$.

Supplementary Fig. 3 illustrates the double sweep transfer characteristics of a MoS$_2$ LH-FET for $V_{ds}$ = 1 V and 100 mV presented in log scale. Notably, the device is completely in the off-state at $V_{gs}$ = - 30 V.

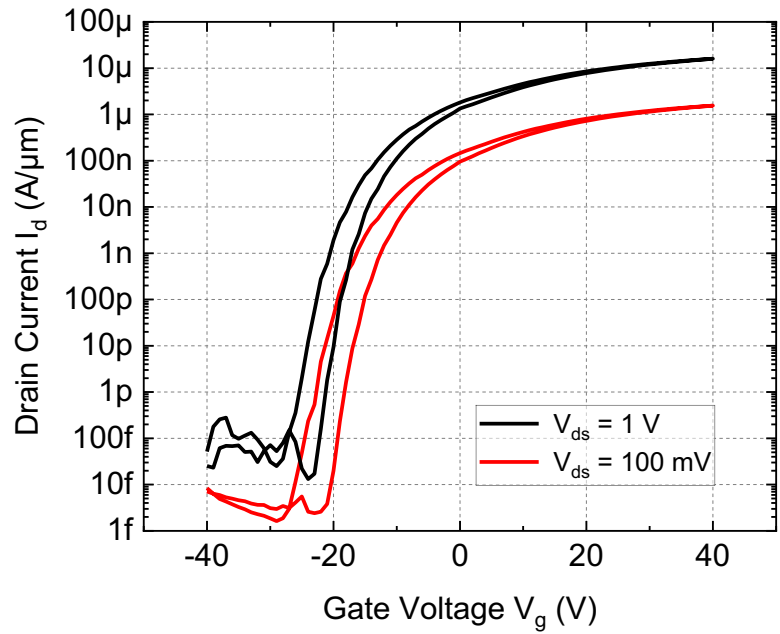

**Supplementary Figure 3**: Transfer curves at $V_{ds}$ = 100 mV and $V_{ds}$ = 1 V of a MoS$_2$ transistor with graphene contacts.

TLM test structures with Ni edge contacted SLG channels were fabricated to assess graphene quality in addition to Raman spectroscopy. Using the TLM method, a contact resistance of $R_c$ = 275 ± 455 Ω µm and a sheet resistance of $R_s$ ~1 kΩ /square were determined (Supplementary Fig. 4).

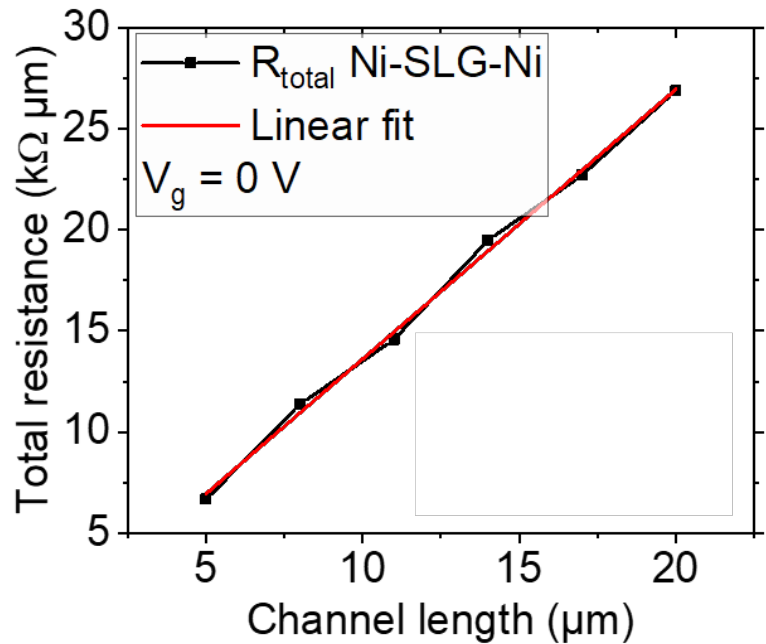

**Supplementary Figure 4:** Total device resistance vs channel length measured by TLM for a Ni edge contacted SLG-FET.

In addition to the LH FETs, Ni edge contacted MoS$_2$ FETs were also fabricated. Using the TLM method, a contact resistance of R$_c$ = 133 ±37 kΩ μm and a sheet resistance of R$_s$ ~34 kΩ /square were determined (Supplementary Fig. 5).

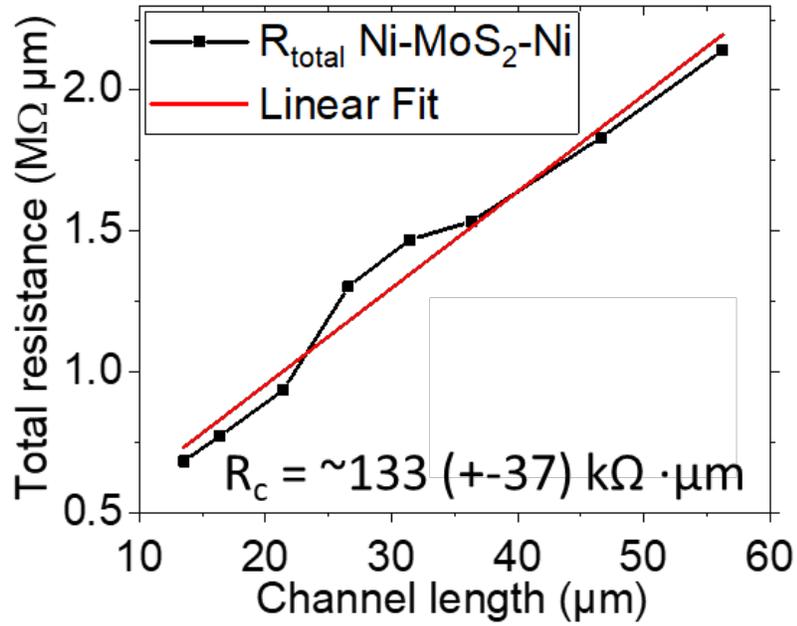

**Supplementary Figure 5**: Total device resistance vs channel length measured by TLM for a Ni edge contacted MoS$_2$ FET.

Supplementary Fig. 6 presents a direct comparison of transfer characteristics for FETs distinguished solely by their contact metals. The results distinctly reveal that devices with pure Ni contacts exhibit significantly higher contact resistance and lower current levels.

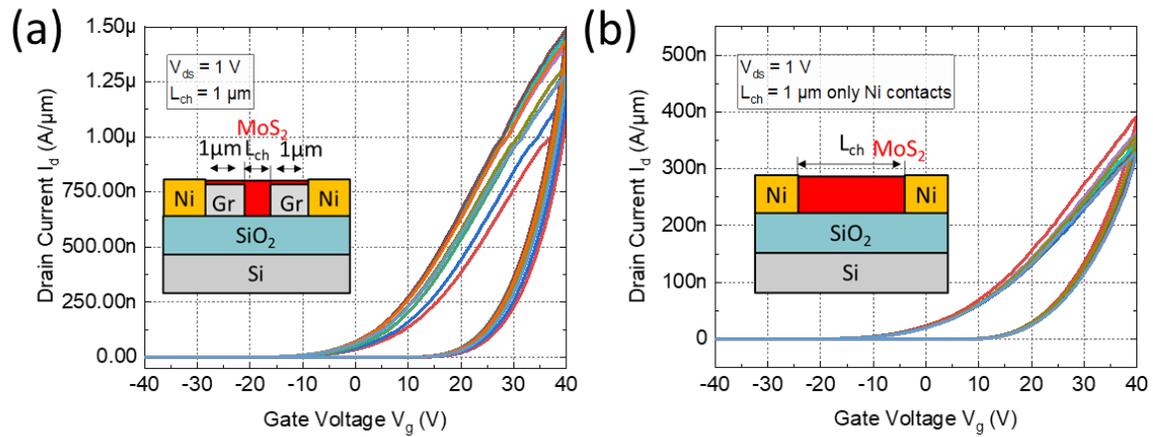

**Supplementary Figure 6:** Comparison of a MoS$_2$ transistor with (a) SLG contacts and (b) only Ni side contacts.

LH-FET (S1) with the same procedure described here in the manuscript were fabricated out of few-layer $MoS_2$ [3] and SLG. A $R_c$ = 13 ±6 kΩ μm and a sheet resistance of $R_s$ ~147 kΩ / □ were determined for the test structures using the TLM method.

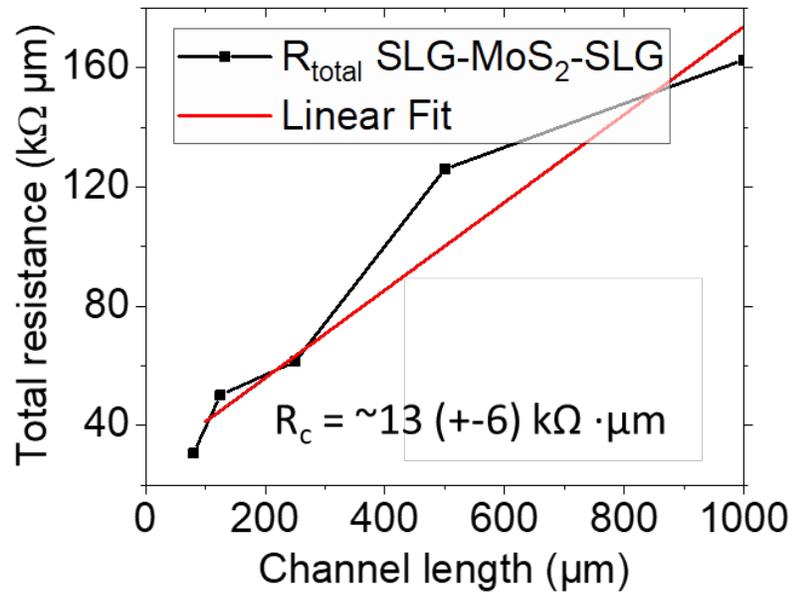

**Supplementary Figure 7**: Total device resistance vs channel length measured by TLM for few-layer $MoS_2$ LH-FET.

The reproducibility of LH-FETs is assessed across 10 different devices for all the 4 channel lengths. Minimal variability is observed among the devices, providing evidence for the scalability of the process.

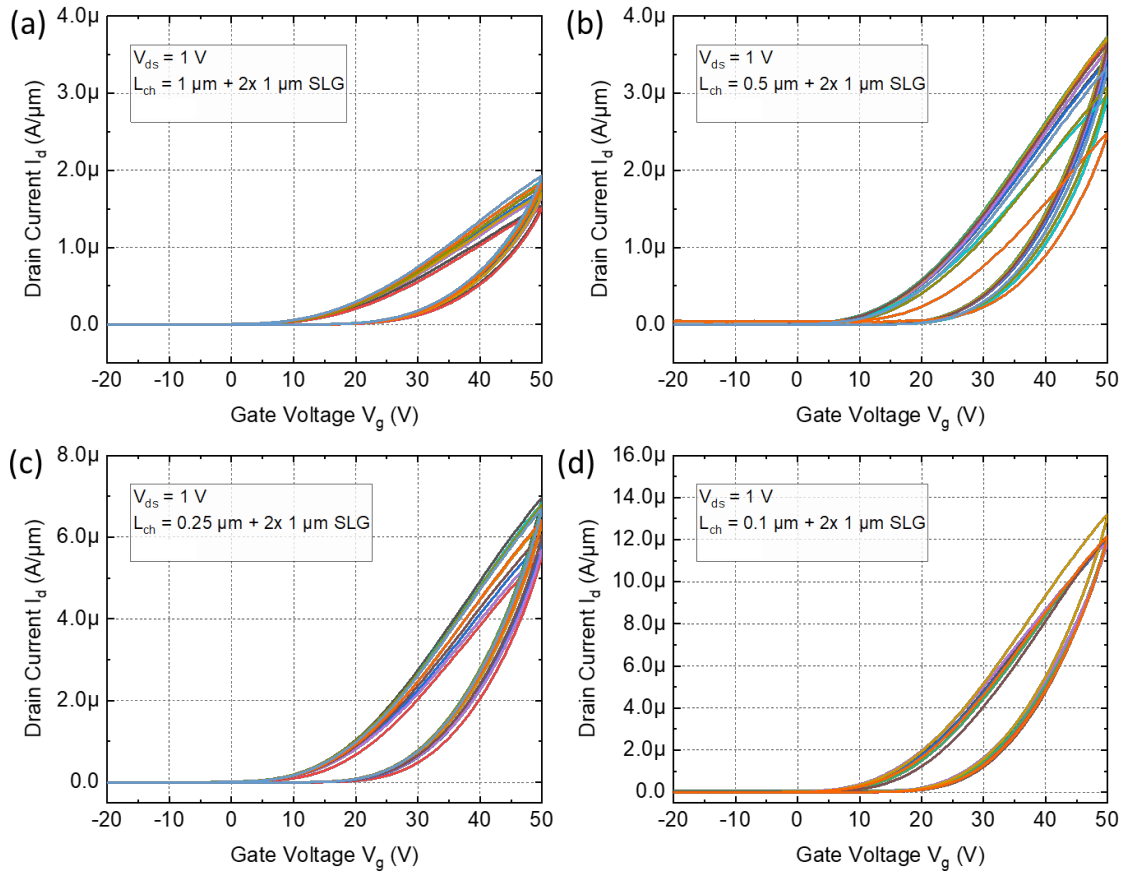

**Supplementary Figure 8**: MoS$_2$ FETs with L$_{ch}$ of (a) 1 μm (b) 0.5 μm, (c) 0.25 μm and (d) 0.1 μm.

Supplementary References